\documentclass[twocolumn]{aastex631}

\usepackage{xspace}

\usepackage{amsmath}
\usepackage{natbib}
\usepackage{enumitem}
\usepackage{caption}
\usepackage{graphicx}
\usepackage[T1]{fontenc}
\usepackage{booktabs}

\DeclareCaptionFormat{cont}{#1 (cont.)#2#3\par}

\newcommand{\halpha}{\ensuremath{\textrm{H}\alpha}\xspace}
\newcommand{\hbeta}{\ensuremath{\textrm{H}\beta}\xspace}
\newcommand{\OIII}{\ensuremath{[\mathrm{O}\textsc{ iii}]}\xspace}
\newcommand{\NII}{\ensuremath{[\mathrm{N}\textsc{ ii}]}\xspace}

\newcommand{\AV}{\ensuremath{A_{\mathrm{V}}}\xspace}

\newcommand{\AVneb}
{\ensuremath{A_{\mathrm{V}, \mathrm{neb}}}\xspace}
\newcommand{\AVstellar}
{\ensuremath{A_{V, \mathrm{stellar}}}\xspace}

\newcommand{\Msun}{\ensuremath{M_{\odot}}\xspace}

\shorttitle{Composite SEDs from MOSDEF Galaxies}
\shortauthors{Lorenz et al.}

\graphicspath{./}

\begin{document}

\title{Stacking and Analyzing $\boldsymbol{z\approx 2}$ MOSDEF Galaxies by Spectral Types:\\Implications for Dust Geometry and Galaxy Evolution}

\author[0000-0002-5337-5856]{Brian Lorenz}
\affiliation{Department of Astronomy, University of California, Berkeley, CA 94720, USA}
\author[0000-0002-7613-9872]{Mariska Kriek}
\affiliation{Leiden Observatory, Leiden University, P.O. Box 9513, 2300 RA Leiden, The Netherlands}
\author[0000-0003-3509-4855]{Alice E. Shapley}
\affiliation{Department of Physics and Astronomy, University of California, Los Angeles, 430 Portola Plaza, Los Angeles, CA 90095, USA}
\author[0000-0003-4792-9119]{Ryan L. Sanders}
\affiliation{Department of Physics and Astronomy, University of Kentucky, 505 Rose Street, Lexington, KY 40506, USA}
\author[0000-0002-2583-5894]{Alison L. Coil}
\affiliation{Center for Astrophysics and Space Sciences, Department of
Physics, University of California, San Diego, 9500 Gilman Dr., La
Jolla, CA 92093, USA}
\author[0000-0001-6755-1315]{Joel Leja}
\affiliation{Department of Astronomy and Astrophysics, The Pennsylvania State University, University Park, PA 16802, USA}
\author[0000-0001-5846-4404]{Bahram Mobasher}
\affiliation{Department of Physics and Astronomy, University of California, Riverside, 900 University Ave., Riverside, CA 92521, USA}
\author[0000-0002-7524-374X]{Erica Nelson}
\affiliation{Department for Astrophysical and Planetary Science, University of Colorado, Boulder, CO 80309, USA}
\author[0000-0002-0108-4176]{Sedona H. Price}
\affiliation{Department of Physics and Astronomy and PITT PACC, University of Pittsburgh, Pittsburgh, PA 15260, USA}
\author[0000-0001-9687-4973]{Naveen A. Reddy}
\affiliation{Department of Physics and Astronomy, University of California, Riverside, 900 University Ave., Riverside, CA 92521, USA}
\author[0000-0003-4852-8958]{Jordan N. Runco}
\affiliation{Department of Physics and Astronomy, University of California, Los Angeles, 430 Portola Plaza, Los Angeles, CA 90095, USA}
\author[0000-0002-1714-1905]{Katherine A. Suess}
\affiliation{Kavli Institute for Particle Astrophysics and Cosmology and Department of Physics, Stanford University, Stanford, CA 94305, USA}
\author[0000-0003-4702-7561]{Irene Shivaei}
\affiliation{Steward Observatory, University of Arizona, Tucson, AZ 85721, USA}
\author[0000-0002-4935-9511]{Brian Siana}
\affiliation{Department of Physics and Astronomy, University of California, Riverside, 900 University Ave., Riverside, CA 92521, USA}
\author[0000-0002-6442-6030]{Daniel R. Weisz}
\affiliation{Department of Astronomy, University of California, Berkeley, CA 94720, USA}

\begin{abstract}
We examine star-formation and dust properties for a sample of 660 galaxies at $1.37\leq z\leq 2.61$ in the MOSDEF survey by dividing them into groups with similarly-shaped spectral energy distributions (SEDs). For each group, we combine the galaxy photometry into a finely-sampled composite SED, and stack their spectra. This method enables the study of more complete galaxy samples, including galaxies with very faint emission lines. We fit these composite SEDs with \texttt{Prospector} to measure the stellar attenuation and SED-based star-formation rates (SFRs). We also derive emission-line properties from the spectral stacks, including Balmer decrements, dust-corrected SFRs, and metallicities. We find that stellar attenuation correlates most strongly with mass, while nebular attenuation correlates strongly with both mass and SFR. Furthermore, the excess of nebular compared to stellar attenuation correlates most strongly with SFR. The highest SFR group has 2\,mag of excess nebular attenuation. Our results are consistent with a model in which star-forming regions become more dusty as galaxy mass increases. To explain the increasing excess nebular attenuation, we require a progressively larger fraction of star formation to occur in highly-obscured regions with increasing SFR. This highly-obscured star formation could occur in dusty clumps or central starbursts. Additionally, as each galaxy group represents a different evolutionary stage, we study their locations on the UVJ and SFR-mass diagrams. As mass increases, metallicity and dust attenuation increase, while sSFR decreases. However, the most massive group moves towards the quiescent region of the UVJ diagram, while showing less obscuration, potentially indicating removal of dust. 
\end{abstract}


\section{Introduction} \label{sec:intro}

In the past two decades we have learned a wealth of information about galaxies at ``cosmic noon'', the peak of star-formation rate (SFR) density in the universe \citep{madau_cosmic_2014, forster_schreiber_star-forming_2020}. Large photometric surveys have provided a wide range of properties, tracing the star-formation history of the universe, how mass builds up in galaxies, and how galaxy morphology changes over time \citep[e.g.,][]{wuyts_smoother_2012}. Simultaneously, spectroscopic surveys give robust SFRs, Balmer decrements, metallicities and chemical compositions, physical conditions of the gas, kinematics, and outflow properties of galaxies over cosmic time \citep[e.g.,][]{price_mosdef_2020, topping_mosdef_2021, sanders_mosdef_2021, weldon_mosdef_2024}. However, there are still many outstanding questions about galaxies during this epoch. In particular, the distribution and impact of dust in these galaxies are still poorly understood.

One of the challenges to answering these open questions is the difficulty in obtaining an unbiased census of spectroscopic properties in distant galaxies. Many crucial measurements, such as metallicity or dust-corrected SFR, require detections of the faint \hbeta line. Unfortunately, \hbeta grows even fainter in dusty galaxies, which also tend to be more massive \citep[e.g.,][]{garn_predicting_2010, price_direct_2014, shapley_mosfire_2022}. While there are a number of spectroscopic surveys at $z=2$ such as MOSDEF \citep{kriek_mosfire_2015}, KBSS \citep{steidel_strong_2014}, and KMOS-3D \citep{wisnioski_kmos3d_2015}, precise measurements of emission-line properties for a complete sample of galaxies, including the most dusty galaxies, has remained a challenge. 

Significantly deeper spectra are required to measure the fainter emission lines for individual galaxies, which are currently being collected with JWST \citep[e.g.,][]{rigby_jwst_2023, finkelstein_ceers_2023, eisenstein_overview_2023, belli_star_2024}. However, there is already a wealth of ground-based data available that enable a complementary approach to study galaxy properties of a complete sample of galaxies. Instead of focusing on individual galaxies, available large samples can be leveraged by stacking spectra and photometry of a number of galaxies to dramatically increase the signal, as demonstrated in \citet{runco_mosdef_2022} for galaxies in the MOSDEF survey. One limitation of these stacking methods is the risk of combining galaxies with different properties or washing out spectral features, and therefore arriving at results that may be difficult to interpret. However, this challenge can be overcome by using photometry to group galaxies with similar spectral types before stacking them. 

Stacking photometry of distant galaxies with similar SED shapes has been demonstrated using medium and broad-band photometric surveys \citep[e.g.,][]{kriek_h_2011, yano_relation_2016, forrest_zfourge_2018, suess_dissecting_2021}. However it is impossible to study metallicity and nebular attenuation with photometry alone. Here we add the power of stacked spectroscopy, combined with stacked photometery based on SED shape, to enable the study of dust and metals across different galaxy types. The dust properties are of particular interest, as they are challenging to measure for individual systems and crucial for interpreting all other measurements. 

In order to create these photometric and spectral stacks, we require a survey of galaxies with deep spectroscopy and deep multi-wavelength photometry. Although stacking does mitigate the uncertainties, we still require deep enough spectra to measure faint emission lines such as \hbeta in the stacks. Finally, a large sample of galaxies is necessary in order to form enough groups of galaxies with similar SED shapes. 

The MOSDEF survey meets all of the above criteria, taking deep spectra of roughly 1500 galaxies that lie in well-observed fields \citep[e.g. CANDELS,][]{grogin_candels_2011, koekemoer_candels_2011}. Furthermore, the recent work of \citet{lorenz_updated_2023} finds that inclination has very little effect on dust attenuation properties. Therefore, we can now freely group the galaxies without worrying about different inclinations. Additionally, modern computing algorithms allow for more sophisticated grouping of galaxies without relying heavily on human judgement. Using these methods, we can more accurately ensure that similar galaxies are being grouped together.

Here, we present the stacked spectra and composite SEDs of 20 groups of galaxies classified by SED shape from the MOSDEF survey. From these data, we derive \halpha SFRs, Balmer decrement dust measurements, O3N2 metallicities, and model them with the stellar population synthesis code \texttt{Prospector} \citep{leja_deriving_2017, johnson_stellar_2021}. In this work, we begin with a description of the details of the MOSDEF dataset and our sub-sample selection (Section \ref{sec:data}). Then we walk through our analysis methods, including the algorithmic grouping of galaxies with similar SED shapes, formation of composite SEDs, stacking of spectra, measurements of emission lines, and fits to the composite SEDs with \texttt{Prospector} (Section \ref{sec:data_analysis}). Next, we present where each of the composite groups lies along the SFR-mass diagram and UVJ diagrams, as well as a summary of the stellar and nebular attenuation properties (Section \ref{sec:results}). We interpret these results in the context of galaxy evolution and possible dust models in Section \ref{sec:discussion}. Finally, we summarize our major findings in Section \ref{sec:summary}. 

Throughout this work we assume a $\Lambda$CDM cosmology with $\Omega_m=0.3$, $\Omega_{\Lambda}=0.7$, and $H_0=70$ km s$^{-1}$ Mpc$^{-1}$. 

\section{Data} \label{sec:data}

\subsection{The MOSDEF Survey} \label{subsec:mosdef_survey}

We analyze galaxy spectra from the MOSFIRE Deep Evolution Field (MOSDEF) Survey \citep{kriek_mosfire_2015}. MOSDEF observed $\sim$$1500$ H-band selected galaxies in the CANDELS fields \citep{grogin_candels_2011, koekemoer_candels_2011} using the MOSFIRE spectrograph \citep{mclean_mosfire_2012} on the Keck I telescope from December 2012 to May 2016. The rest-frame optical spectra are of moderate resolution ($R=3000-3650$) and observed in the $Y, J, H,$ and $K$ atmospheric windows depending on redshift. MOSDEF observed targets in three redshift regimes ($1.37\leq z\leq 1.70, 2.09\leq z\leq 2.61,$ and $2.95\leq z \leq 3.60$). We refer to \citet{kriek_mosfire_2015} for additional information about the MOSDEF survey, including sample selection, observations, reduction, and ensemble properties.

We also use the v4.1 photometric catalogs from the 3D-HST survey \citep{brammer_3d-hst_2012, skelton_3d-hst_2014, momcheva_3d-hst_2016}. We adopt the stellar population properties derived from fitting the emission-line corrected SEDs using the FAST fitting code \citep{kriek_ultra-deep_2009}, which assumes flexible stellar population models \citep[FSPS,][]{conroy_propagation_2009, conroy_propagation_2010}, MOSFIRE spectroscopic redshifts, a \citet{chabrier_galactic_2003} stellar initial mass function (IMF), the dust attenuation curve from \citet{calzetti_dust_2000}, delayed exponentially declining star formation histories, and solar metallicity. These fits yielded measurements of stellar mass and dust attenuation \AV for individual galaxies, and have been used in a number of MOSDEF studies \citep{sanders_mosdef_2018, sanders_mosdef_2020, sanders_mosdef_2021, shapley_mosdef_2015, shapley_mosdef_2019, shapley_mosfire_2022, runco_mosdef_2021, runco_mosdef_2022}.

Metallicities and dust-corrected SFRs for individual galaxies were measured through fits to prominent emission lines \citep{reddy_mosdef_2015}, including \halpha and \hbeta. The best-fit stellar population models are used to correct the Balmer lines for underlying absorption. For the galaxies with detected \halpha but no strong detections in \hbeta, we do not have a Balmer decrement dust measurement. For these cases, we calculate an \halpha SFR applying a dust-correction according to the FAST \AV, assuming that the nebular \AV is a factor of 2 larger than the stellar \AV, as proposed in \citet{calzetti_dust_2000}. We take this simple correction since the individual galaxy SFR values are not essential to the conclusions in this work. Given our results based on the stacks in Section \ref{subsec:dust_results}, Figure \ref{fig:av_compare}, we indeed find that this factor is applicable for our sample \citep[see also][for a different conversion method between E(B-V)$_\mathrm{gas}$ and E(B-V)$_\mathrm{stars}$]{sanders_mosdef_2021}. For galaxies without \halpha and \hbeta detections, we do not derive an individual SFR based on the Balmer emission lines.

\subsection{Sample Selection} \label{subsec:sample_select}
In this work, we aim to measure SFRs and dust properties of MOSDEF galaxies. To do so, we aim to create highly-sampled composite SEDs and deep, stacked spectra by grouping galaxies by spectral type. Then, we can measure their emission-line properties. In order to accurately dust-correct the galaxies to measure SFR, we need coverage of the \halpha and \hbeta lines. Therefore, we only use the MOSDEF galaxies in the lower and middle redshift regimes ($1.37\leq z\leq 1.70, 2.09\leq z\leq 2.61$) for which \halpha falls in a band of atmospheric transmission. We then restrict the remaining 1057 galaxies by the following criteria to ensure that we can make robust measurements of the emission properties of the stacked spectra.

First, the 138 IR, X-ray, and optical AGN above the \citet{kewley_cosmic_2013} line identified in the MOSDEF survey are removed \citep{coil_mosdef_2015, azadi_mosdef_2017}, ensuring that the emission features are originating from star formation for all galaxies. Then, we remove another 106 galaxies that do not have high-quality redshift measurements from MOSDEF, and consequently would not be able to be used in spectral stacks since we could not align their emission features. Next, we remove 99 serendipitous detections, ensuring that the galaxies all meet the MOSDEF targeting criteria and are aligned in the middle of slits. We then remove 44 galaxies that do not have coverage of both \halpha and \hbeta emission lines. Finally, we later remove 10 galaxies that do not fit well in any group (see Section \ref{subsec:form_composite_sed}) leaving 660 galaxies. In this sample, 408 galaxies have individually $3\sigma$ detected \halpha and \hbeta, 196 have only $3\sigma$ detected \halpha, and 56 have neither line detected at $3\sigma$. For these 56 galaxies, the redshifts were derived from multiple emission lines with $2\sigma$ detections or robust absorption lines.

\section{Data Analysis} \label{sec:data_analysis}

In this work, we aim to stack galaxies with similar spectral shapes to obtain high-quality well-sampled SEDs and deep spectra to measure emission lines that are too faint to detect in individual galaxy spectra. Therefore, we first employ an algorithm to separate galaxies into groups with similar SED shapes. Then, we use individual galaxy SED points to form a highly-sampled composite SED, and stack spectra for measurements of emission lines. With the stacked spectra, the dramatic increase in signal-to-noise ratio compared to an individual galaxy spectrum allows for measurement of the \hbeta line for nearly all groups. Detected \hbeta provides accurate nebular dust properties, and thus also more accurate SFR measurements. In this section, we walk through the steps to obtain these measurements. 

\subsection{Grouping Galaxies by SED Shape} \label{subsec:galaxy_shapes}

To make composite SEDs, we must form groups of galaxies with similar properties. We use a similar approach to how composite SEDs have been formed in other studies \citep[e.g.,][]{kriek_h_2011, forrest_zfourge_2018, suess_dissecting_2021}, with two notable differences. First, our sample has spectroscopic redshifts, an improvement over the photometric redshifts that were used in prior works. Second, we use spectral clustering to group the galaxies, a more objective method than what has been previously employed. 

In order to group the galaxies by SED shape, we first must employ a method to compare galaxy SED shapes. All of the galaxies are observed with the same filters, but they are at slightly different redshifts, and therefore the filters cannot be compared directly. Instead, we de-redshift each galaxy SED to rest-frame and linearly interpolate each observed SED into 20 equally spaced (in log(wavelength)) mock filters from $1300$\,\AA\ to $20000$\,\AA, so every filter has a width of $\approx 0.062$ dex. Specifically, we assume a constant transmission within each filter, and that there are no gaps between the filters. For example, the first filter is a top hat that covers $1300-1501$\AA, and the second top hat filter is from $1501-1733$\AA. To measure the galaxy SED in a mock filter, we make a local quadratic fit to the SED and then observe it with the mock filter. We make the quadratic fit using all photometric data points that are $\pm0.25$ dex from the center of the filter in log(wavelength). Then, we take the average value of the quadratic within the filter range to infer the measured mock SED point. Repeating this technique across the entire galaxy SED, we obtain measurements for each galaxy in the same 20 mock rest-frame filters. 

Next, we use these interpolated fluxes to compare the SED shapes of each galaxy in our sample. We perform this comparison with the following cross-correlation equations from \citet{kriek_h_2011}, which normalize the SEDs to each other and then compute a metric for how similar their points are, in a $\chi^2$ manner. For galaxies 1 and 2, where $f_{\lambda,1}$ and $f_{\lambda,2}$ represent the interpolated flux values for the SEDs of galaxies 1 and 2 respectively, we compute the normalization factor $a_{12}$ and similarity metric $b_{12}$ as

\begin{equation}\label{eq:a12}
    a_{12} = \frac{\sum(f_{\lambda,1} f_{\lambda,2})}{\sum(f_{\lambda,2})^2}
\end{equation}
\begin{equation}\label{eq:b12}
    b_{12} = 1-\sqrt{\frac{\sum(f_{\lambda,1} - a_{12}f_{\lambda,2})^2}{\sum(f_{\lambda,1})^2}}.
\end{equation}
$b_{12}$ is a value from 0 to 1, where $b=0$ represents perfect anti-correlation and $b=1$ represents identical shapes. 

To form our groups of galaxies with similar SED shapes, we employ a spectral clustering algorithm \citep{von_luxburg_tutorial_2007, pedregosa_scikit-learn_2011, damle_simple_2019}. By algorithmically clustering SEDs, we remove any sort of human bias as to which types of galaxies should be clustered first, or which galaxies should be excluded from clusters. Instead, the algorithm optimizes the grouping such that SED shapes within each group are as similar as possible, with each group being as distinct from one another as possible.

We apply this correlation measure to all of the mock galaxy SEDs. We correlate each galaxy with every other one, storing the resulting $b_{12}$ values in a similarity matrix. This similarity matrix forms the basis of the spectral clustering algorithm that we employ. First, we plot the eigenvalues of the matrix to determine the appropriate number of clusters. Each eigenvalue represents a principal component of similarity, and thus we want to select a number that captures as many high-signal principle components as possible. We search the eigenvalue graph for the point where the derivative of the eigenvalues falls below $-0.1$, so the eigenvalues are not changing much from one point to the next, and therefore do not have significant signal in their components. For this work, the derivative of the eigenvalues falls below $-0.1$ when there are 20 clusters. We then run the spectral clustering algorithm. This algorithm uses the specified 20 eigenvalues from the similarity matrix, which contain the highest-signal components that describe the galaxies' shapes, to embed the data into a 20-dimensional space instead of the original 660 dimensions. In this lower-dimensional space, it applies a k-means clustering to form the 20 groups. This grouping is completely model-independent --- we form the groups solely on their observed fluxes and spectroscopic redshifts, and do not assume anything about the intrinsic properties of the galaxies. 

\subsection{Forming Composite SEDs}
\label{subsec:form_composite_sed}

By combining the SEDs of all of the individual galaxies within a group, we create a highly-sampled composite SED from which we can measure group properties. To form the composite SED, we first normalize all of the galaxies using $a_{12}$ from equation \ref{eq:a12} so that their SEDs are scaled to the same level. Arbitrarily, we pick the galaxy in the group with the lowest HST catalog ID number to be the target galaxy, then normalize all other SEDs to this target. Then we sort all flux measurements of the galaxies by rest wavelength, regardless of which galaxy the measurements originally came from. We combine the first $k$ photometric datapoints into the first composite datapoint, where $k=\lfloor\frac{n}{3}\rfloor$ and $n$ is the number of galaxies in that cluster. The flux of the composite point is computed as the mean of the $k$ fluxes contributing to it, with uncertainties computed as the 16th and 84th percentile of contributing points. Similarly, the wavelength is the mean of the $k$ contributing wavelengths. Then, this process is repeated for the next $k$ points. In the final bin, if the total number of points is not perfectly divisible by $k$, any remaining points are grouped into the previous composite point. 

Then the composite SED filters are computed. For each composite SED point, the filter is a combination of the $k$ filters that contribute to it. The response curve of each contributing filter is de-redshifted to rest frame, then linearly interpolated at very high resolution to the same wavelength scale. The response curve for the filter for the composite SED point is then the mean of these interpolated response curves. 

As a sanity check, we compare the individual SEDs to their group composite using the cross-correlation described in Section \ref{subsec:galaxy_shapes}. A few galaxies are placed into groups that they don't match well, since every galaxy is forced to be placed somewhere. To combat this issue, any galaxies with similarities less than 0.8 are removed from their groups (10 galaxies), and the composites are re-formed without them. 

\subsection{Spectral Stacking} \label{subsec:spectral_stacking}

In addition to the composite SEDs, we also combine the individual MOSDEF spectra to form composite spectra, from which we can measure detailed emission properties for all groups. 

Spectra are stacked in the same manner as \citet{lorenz_updated_2023}. In summary, spectral luminosities are stacked in the rest-frame with no normalization. Sky lines are masked out, and each spectrum is interpolated to 0.5\AA\ per pixel. Each pixel of the stacked spectrum is a median of the non-masked pixels contributing to it (refer to \citet{lorenz_updated_2023} for a discussion of mean vs. median stacking). FAST stellar continuum models of each of the galaxies are similarly stacked, providing a model continuum and underlying absorption for emission line fitting. Since the spectra are median-stacked, we also measure the mass of the group as a median of the masses of all contributing galaxies. 

The composite SEDs, stacked spectra, and summary of properties for each group are presented in Figure \ref{fig:composite_seds_page0}. The number of individual galaxies with 3$\sigma$ \halpha and \hbeta detections (and therefore SFRs) is shown in the middle panel, and the number of galaxies with strong detections in all four BPT lines is shown in the rightmost panel. For most of the groups, more than half of the individual galaxies cannot be confidently placed on the BPT diagram, but the stacked spectrum has $3\sigma$ detections in all four lines (see Section \ref{subsec:emission_fit}). 

\begin{figure*}[tp]
\vglue -5pt
\centering
\includegraphics[width=\textwidth]{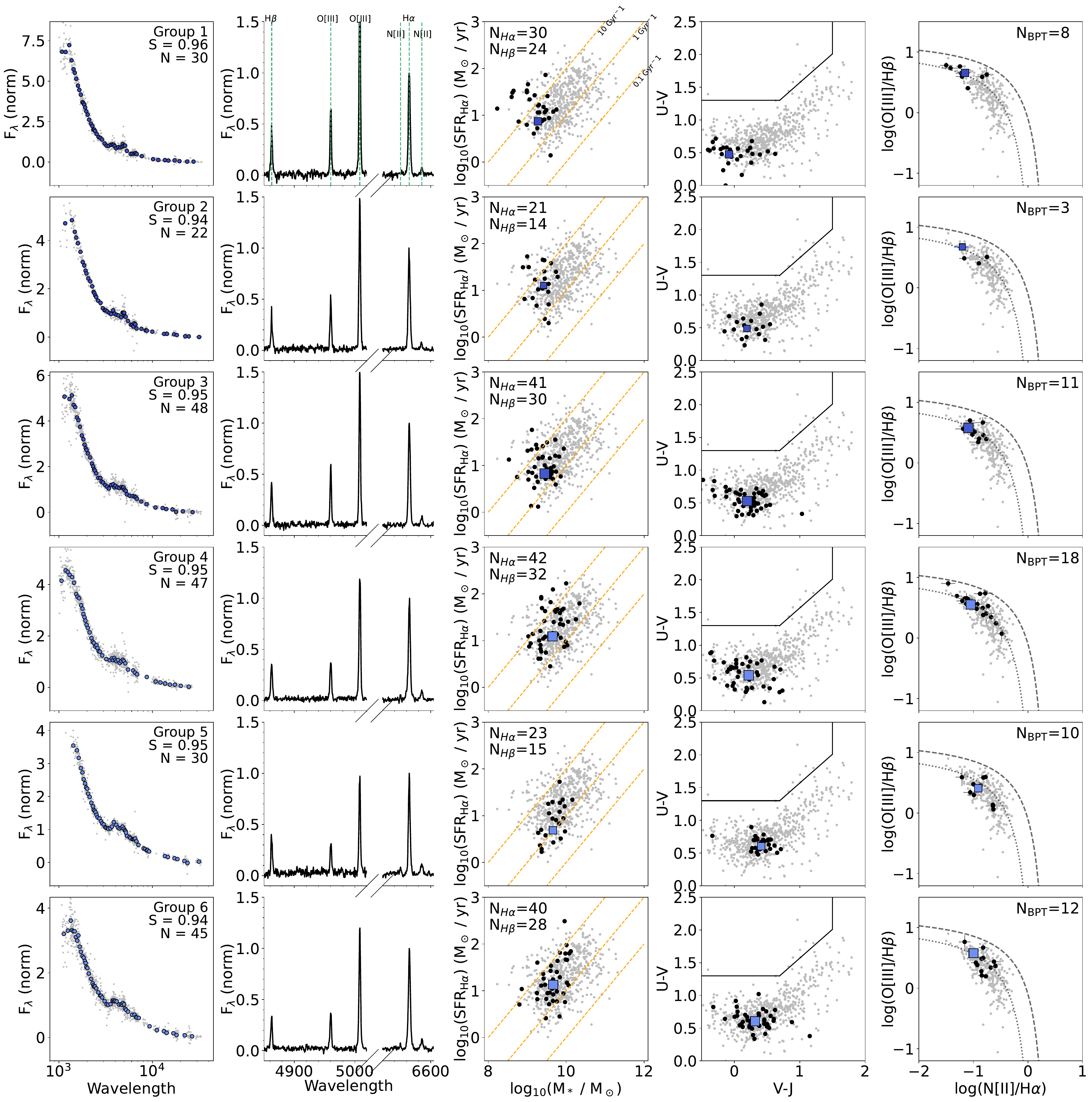}
\caption{
Overview of properties for the 20 galaxy groups, sorted by the median mass of the galaxies in the group. For each group, from left to right we show: a) composite SED points (color) and all contributing galaxy SEDs (gray). The composite SEDs are normalized for viewing such that flux is 1 at 5000\AA. We also show the number of galaxies in the group (N) and average similarity (S) between the individual galaxy SEDs and composite. b) stacked spectrum, zoomed on emission line regions. For viewing, the spectra are normalized such that peak \halpha flux is 1. c) SFR vs mass, with lines of constant sSFR shown (yellow). The number of galaxies in the group with \halpha and \hbeta detections is listed in the top left. d) UVJ diagram, showing the top-left enclosed region for quiescent galaxies. e) BPT diagram, with the \citet{kewley_cosmic_2013} line (black) and \citet{shapley_mosdef_2015} MOSDEF fit (gray) shown. In the top-right we list the number of galaxies in the group with 3$\sigma$ detections on all four emission lines. For c), d), and e), we show the entire sample used in this work (grey) with reliable measurements, the galaxies in a particular group (black), and the measurement from the composite SED and stacked spectrum (colored symbol). Points representing the group in panels c), d), and e) scale in size with the number of galaxies in the group and are colored by mass (see Figure \ref{fig:ssfr_compare_targetmass}). These points will remain consistent throughout the paper. 
}
\label{fig:composite_seds_page0}
\end{figure*}

\begin{figure*}[tp]
\vglue -5pt
\ContinuedFloat
\captionsetup{list=off,format=cont}
\centering
\includegraphics[width=\textwidth]{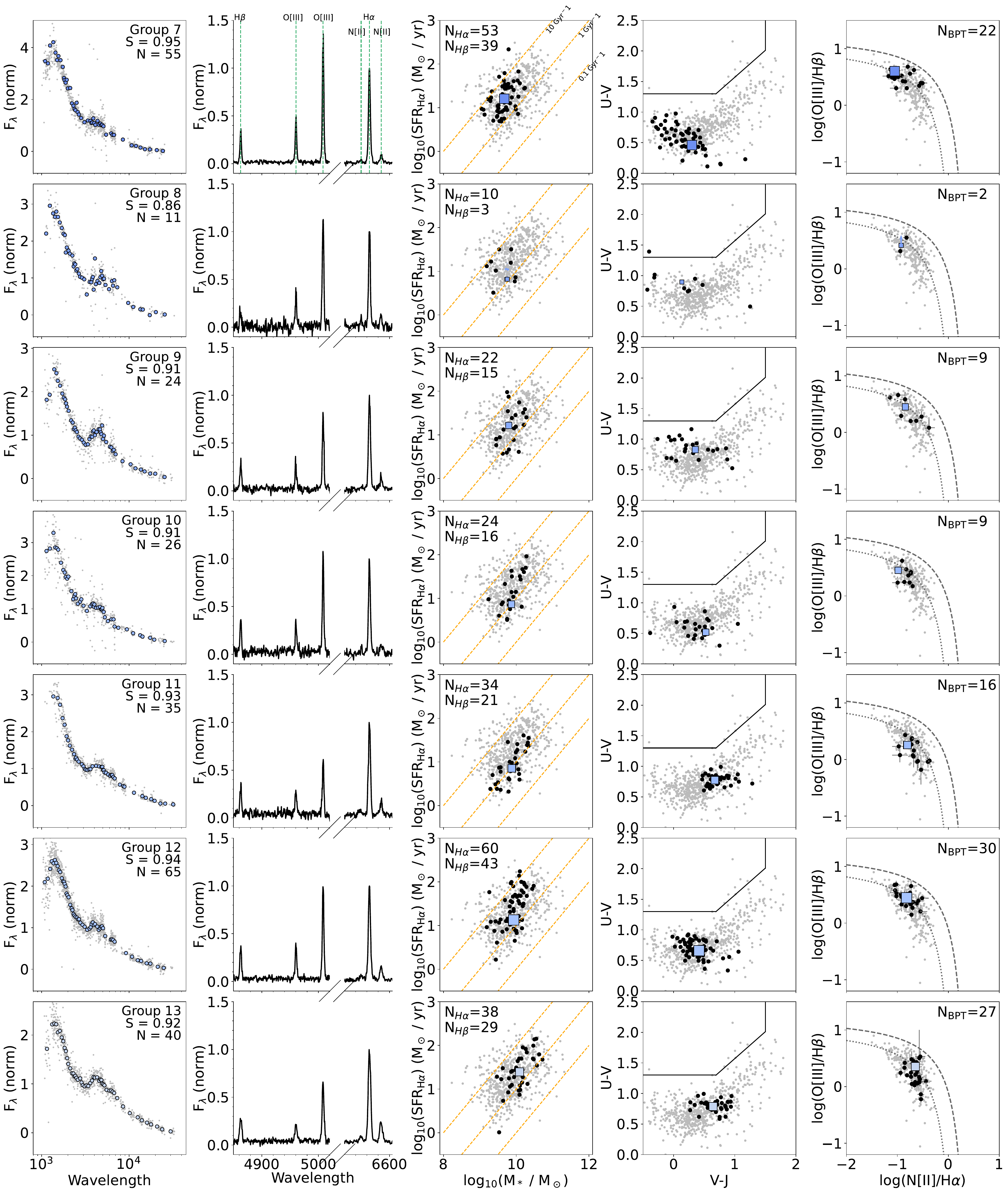}
\caption{
The next 7 groups, sorted by sSFR. Note that group 8 has lower limits on its SFR measurement due to noisy \hbeta. 
}
\label{fig:composite_seds_page1}
\end{figure*}

\begin{figure*}[tp]
\vglue -5pt
\ContinuedFloat
\captionsetup{list=off,format=cont}
\centering
\includegraphics[width=\textwidth]{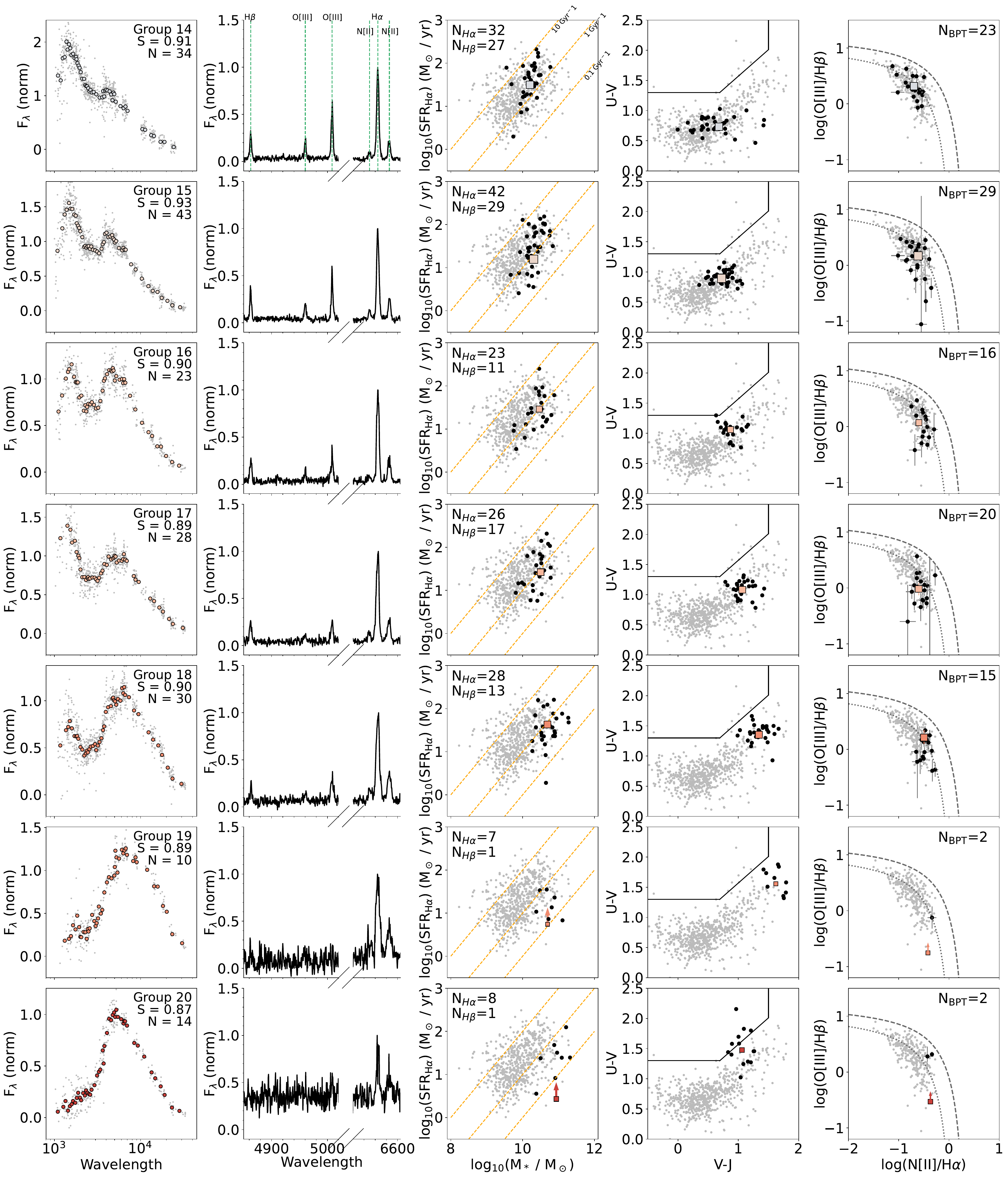}
\caption{
The final 7 groups, sorted by sSFR. Note that groups 19 and 20 have lower limits on their SFR measurements, and their \hbeta lines are not perceptible above the noise. 
}
\label{fig:composite_seds_page2}
\end{figure*}

\subsection{Emission Fitting} \label{subsec:emission_fit}
We fit 6 prominent emission lines in each of the stacked spectra: \hbeta $\lambda4863$, \OIII $\lambda4960$ and $\lambda5008$, \halpha $\lambda6565$, and \NII $\lambda6550$ and $\lambda6585$. We simultaneously fit for redshift, velocity dispersion, and the flux for each line. The continuum is modeled by stacking the FAST SED fits to the individual galaxies using the same scaling factors as for the composite SED. Then, we take this modelled continuum, which includes Balmer absorption, and scale it to match the continuum of the spectrum in the region of the emission lines. All of the line fluxes are free in the fits, but we assume the same redshift and velocity dispersion (in km/s) for all lines. Uncertainties are estimated through bootstrapping. Spectra are stacked again with a random subset of $n$ galaxies, where $n$ is the number of galaxies in the group. Repeating this step 100 times, we fit the emission lines from the 100 bootstrapped spectral stacks. For each line, we determine the uncertainties on the flux, line ratios, and velocity dispersion from the 16th and 84th percentile of the distribution of measurements for each line. Thus, we have line ratio measurements and uncertainties for $\frac{\halpha}{\hbeta}$, $\frac{\NII\lambda \mathrm{6585}}{\halpha}$, $\frac{\OIII\lambda \mathrm{5008}}{\hbeta}$, and O3N2 (see Section \ref{subsec:metallicity_measure}).

\subsection{SFR Calculation}
\label{subsec:sfr_calc}
We measure dust-corrected \halpha SFRs for each of the composite SED groups. 

First, we determine which of the groups do not have a strong enough \hbeta detection in the stacked spectrum to measure an accurate Balmer decrement. We require at least a $3\sigma$ detection of \hbeta, where the uncertainty is determined from the emission fits to the bootstrapped spectra. Three of the groups do not have a sufficiently detected \hbeta line. For these three groups, we place a $2\sigma$ lower-limit on their Balmer decrement using the 2.5th percentile of the bootstrapped Balmer decrement distribution. When this lower-limit is less than the theoretical minimum of 2.86, we assume the lower-limit is 2.86. The dust-correction calculation is then carried out with these lower-limit Balmer decrements, placing $2\sigma$ lower-limits on the SFRs for these three groups. 

To compute SFR, we take the measured \halpha luminosity from the stacked spectra, and perform a dust-correction. The Balmer decrements are converted to \AVneb assuming a \citet{cardelli_relationship_1989} extinction curve with 
\begin{equation}
    \AVneb = R'_V\times 2.32\times \log_{10}\left(\frac{\halpha / \hbeta}{2.86}\right),
\end{equation}
assuming $R'_V = 3.1$. We then calculate the nebular attenuation at the wavelength of \halpha using the \citet{cardelli_relationship_1989} extinction curve (consistent with the findings of \citet{reddy_mosdef_2020} for MOSDEF galaxies), and use this value to compute the dust-corrected \halpha luminosity. Next, we convert the dust-corrected luminosities to SFRs using a conversion factor from \citet{reddy_hduv_2018}, as described in \citet{shapley_jwstnirspec_2023}, which is similar to those presented in \citet{hao_dust-corrected_2011}: 
\begin{equation}
    \mathrm{SFR} = L\left(\halpha\right) \times 10^{-41.37} \frac{\rm{\Msun} \mathrm{yr}^{-1}}{\mathrm{erg \ s}^{-1}}.
\end{equation} 
Finally, since the spectra are median-stacked, sSFRs are calculated by dividing the computed SFR by the median mass in each group. 

\begin{figure*}[tp]
\vglue -5pt
\centering
\includegraphics[width=\textwidth]{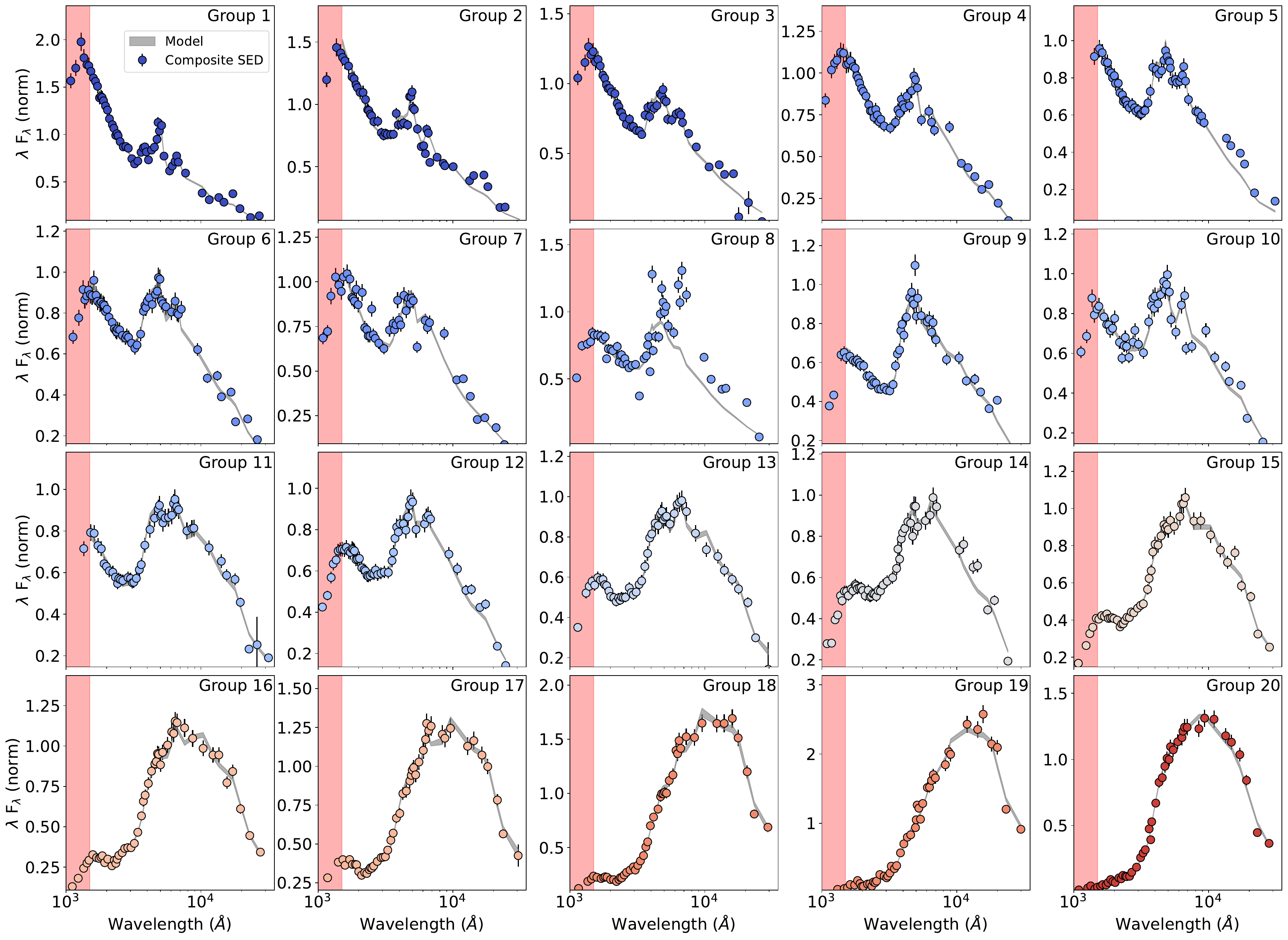}
\caption{\texttt{Prospector} fits to all 20 composite SED groups, ordered and colored by stellar mass (see Figure \ref{fig:mass_sfr_uvj_bpt}). The \texttt{Prospector} 16th to 84th percentile fit is indicated by the shaded gray region. The regions left of $1500$\AA\ (red) are not included in the fits due to Lyman-$\alpha$ forest absorption.
}
\label{fig:prospector_fits}
\end{figure*}

\subsection{Metallicity Measurement} \label{subsec:metallicity_measure}
To measure metallicities for the galaxy groups with the measured emission lines, we use the O3N2 calibration. The O3N2 ratio is computed as 

\begin{equation}
    \mathrm{O3N2} = \log_{10}\left(\frac{\mathrm{\OIII \lambda5008\AA / \hbeta}}{\mathrm{\NII \lambda6585\AA / \halpha}}\right),
\end{equation}
which is a robust measure of metallicity due to the multiple strong-line, dust-independent ratios \citep[e.g.,][]{liu_metallicities_2008, steidel_strong_2014, sanders_mosdef_2015}. Then, we turn this ratio into a metallicity using metallicity measurements of low-redshift analogs of high-redshift galaxies from \citet{bian_direct_2018}. The conversion is

\begin{equation}
    12+\log_{10}\left(O/H\right) = 8.97-0.39\times \mathrm{O3N2}.
\end{equation}

For the three groups that do not have $3\sigma$ detections in \hbeta, we place $2\sigma$ lower-limits on their O3N2 measurement, and therefore $2\sigma$ upper-limits on their metallicity, by using the 2.5th percentile \hbeta luminosity from the distribution of measurements of the bootstrapped spectra. 

\subsection{Prospector SED Fitting} \label{subsec:prospector} In order to measure SED-based SFRs and dust properties, each of the composite SEDs are fit with the SED-fitting code \texttt{Prospector} \citep{leja_deriving_2017, johnson_stellar_2021}, which uses the FSPS stellar population synthesis code \citep{conroy_propagation_2009}, the MILES stellar library \citep{falcon-barroso_updated_2011}, and the MIST isochrones \citep{choi_mesa_2016, dotter_mesa_2016}. Each group is fit at the median redshift of all of its constituent galaxies, and we use the composite filter curves when fitting the composite SED. 

We assume a parametric delayed-tau star-formation history (SFH), which should be reasonable for a composite SED since individual starbursts are likely washed out. We also assume the \citet{kriek_dust_2013} attenuation curve (\texttt{dust\_type} = 4) with a single dust screen, which includes a relationship between the slope of the attenuation curve and the UV dust bump. Finally, we assume the default \citep{kroupa_variation_2001} IMF. The composite SEDs are fit with the following free parameters and priors: 
\begin{itemize}
    \itemsep0em 
    \item Stellar Mass $\left(10^{8}\leq M/\rm{M}_\odot\leq 10^{13}\right)$
    \item Age $\left(0\leq \mathrm{age}\ \left(\mathrm{Gyr}\right)\leq \left(1+\mathrm{Universe\ age\ at\ group\ } z \right)\right)$
    \item Stellar Metallicity $\left(-1.4\leq \log(Z / \mathrm{Z}_\odot) \leq 0.19\right)$
    \item Stellar Dust Optical Depth $\left(0\leq \mathrm{\texttt{dust2}} \leq 4\right)$
    \item Dust Attenuation Curve Slope offset with respect to the \citet{calzetti_dust_2000} law, as parameterized by \citet{noll_analysis_2009}. $\left(-1 \leq \mathrm{\texttt{dust\_index}} \leq 0.4\right)$
    \item Parametric SFH Exponent $\left(0.1\leq \tau \left(\mathrm{Gyr}^{-1}\right) \leq 10\right)$.
\end{itemize} 
The parameter space is explored using dynamic nested sampling with \texttt{dynesty} \citep{speagle_dynesty_2020}. We do not include any SED points bluer than 1500\AA\ due to galaxies in each group being at slightly different redshifts, and therefore having different amounts of intergalactic Lyman-$\alpha$ absorption that could not easily be modelled.  

For each composite SED, we take the 1000 highest-weighted fits from the dynesty sampling, and compute the weighted-median measurement (50th percentile) and the $1\sigma$ uncertainties (16th and 84th percentiles) for each property. \texttt{Prospector} fits are shown in Figure \ref{fig:prospector_fits}, with the 16th and 84th percentiles setting the boundary of the grey regions. The corresponding stellar population properties are listed in Table \ref{tab:sed_table}.

\begin{figure}[tp]
\vglue -5pt
\centering
\includegraphics[width=0.45\textwidth]{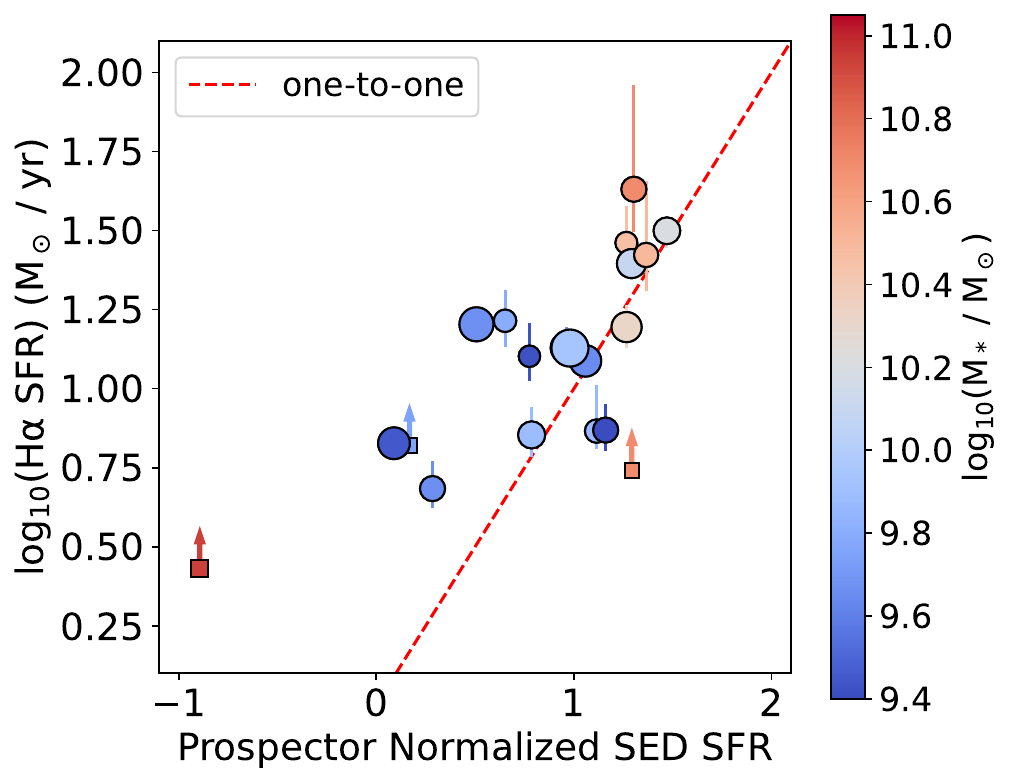}
\caption{Prospector SFR derived from composite SEDs vs. Balmer decrement-corrected \halpha SFR from the stacked spectra. Point size scales with the number of galaxies in the group, squares denote the three groups for which we have insignificant \hbeta measurements, and points are colored by stellar mass.  A one-to-one line is shown in red, suggesting reasonable agreement in the SFR predicted by the SED and measured by dust-corrected \halpha. Group 20 has a much higher \halpha SFR than SED-based SFR, potentially indicating has non-stellar contribution to \halpha.
}
\label{fig:ssfr_compare_targetmass}
\end{figure}

The SFR measurement from \texttt{Prospector} is derived for the composite SED, which is inherently scaled to the target galaxy of the group (see Section \ref{subsec:form_composite_sed}) and shifted to the median redshift of the group. To make the SFR measurement representative of the full group and enable comparison to the SFRs derived from the stacked spectra, we must correct for the arbitrary SED normalization to the target galaxy. To measure a specific SFR, we divide the \texttt{Prospector} SFR by the measured \texttt{Prospector} mass of the group, giving a direct measurement of sSFR from \texttt{Prospector}. Finally, we multiply this sSFR by the median mass of the group to measure SFR, since the spectra are median-stacked. We find that the SFRs predicted from the SED fits are in reasonable agreement with the \halpha SFRs computed from the stacked spectra, as shown in Figure \ref{fig:ssfr_compare_targetmass}. We find that group 20 has a much higher \halpha SFR than SED SFR. This offset is possibly due to AGN contribution to \halpha driving up the measured SFR (see Section \ref{subsec:evolution}). 

Our results are consistent with \citet{shivaei_mosdef_2016}, finding that broad UV-to-NIR SEDs fit with delayed-tau star formation histories are in agreement with the dust-corrected \halpha SFRs from spectra, though we probe even lower stellar masses. This agreement instills confidence in the ability to predict accurate SFRs for high-redshift galaxies that do not have such detailed spectral measurements, although the scatter is still large. See \citet{leja_deriving_2017} for an assessment of how well \texttt{Prospector} can predict \halpha luminosities from photometry.

\begin{figure*}[tp]
\vglue -5pt
\centering
\includegraphics[width=0.95\textwidth]{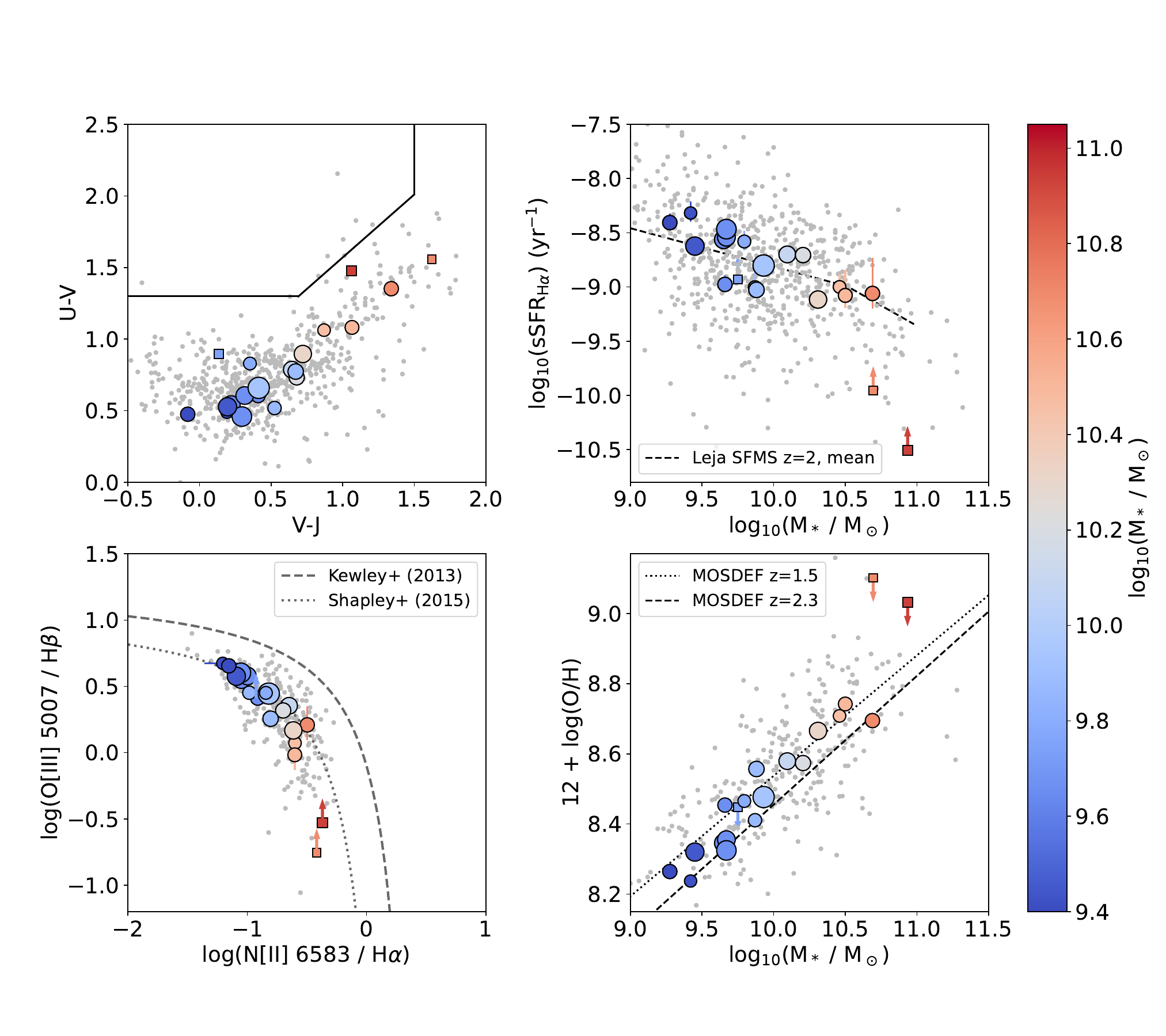}
\vspace{-2\baselineskip}
\caption{Overview of basic properties of the galaxy groups (color, with the same size and symbols as Figure \ref{fig:ssfr_compare_targetmass}) compared to the individual galaxies (gray). Top left: UVJ diagram. Galaxies follow the expected mass trend, with mass increasing towards the upper-right, but the highest-mass group is lower in $V-J$ and moving towards the quiescent box. Top right: sSFR vs stellar mass, with SFR derived from \halpha. Individual galaxies are only shown if the have a $3\sigma$ detection in \halpha. Our trend is in strong agreement with the trends found by \citet{leja_new_2021} using SED SFRs from 3D-HST. Bottom left: BPT diagram, along with the \citet{kewley_cosmic_2013} line (black) and \citet{shapley_mosdef_2015} fit (gray) to MOSDEF galaxies. Our composite SEDs are in strong agreement with this fit, and they follow the expected trend with mass increasing towards the lower-right. Bottom right: Metallicity vs mass, with a comparison to the MOSDEF $z=1.5$ relation from \citet{topping_mosdef_2021} (dotted line) and MOSDEF $z=2.3$ relation from \citet{sanders_mosdef_2018} (dashed line) for the individual MOSDEF galaxies. Our sample, with mean $z=1.9$, falls between these lines. In the BPT and mass-metallicity diagrams, we only show individual galaxies with $3\sigma$ detections in all four lines.
}
\label{fig:mass_sfr_uvj_bpt}
\end{figure*}

\begin{figure}[]
\vglue -5pt
\centering
\includegraphics[width=0.45\textwidth]{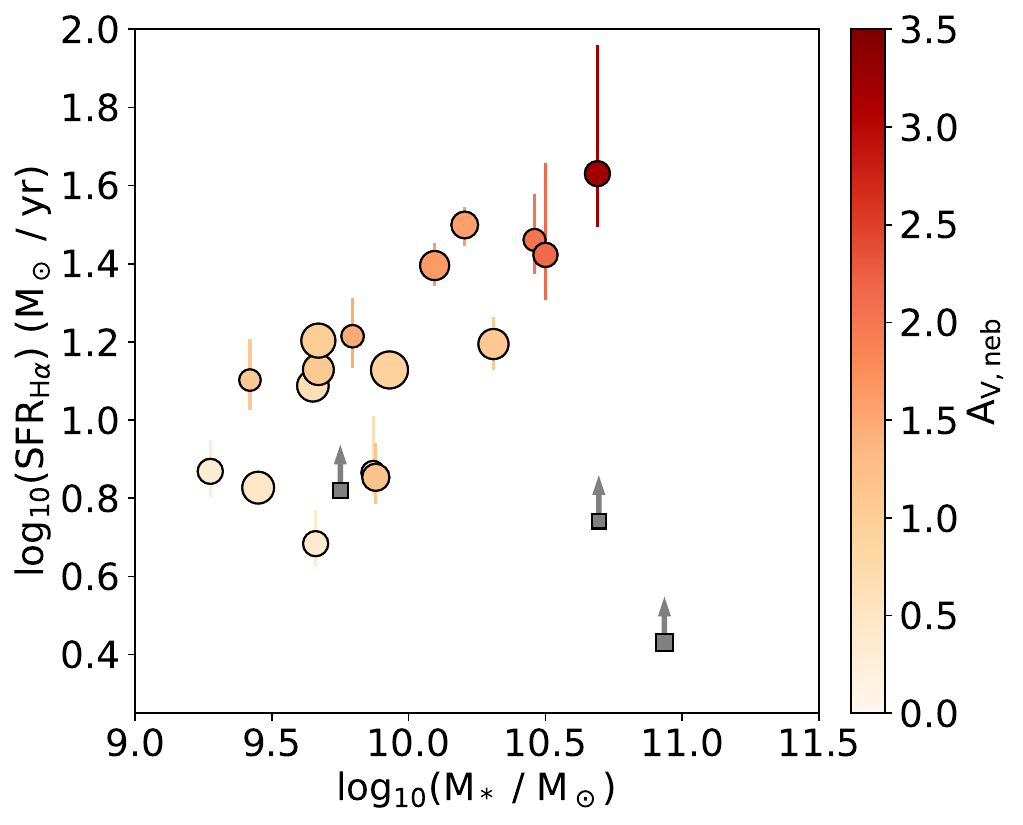}
\caption{SFR vs Mass for the composite groups, color-coded by nebular attenuation. Point sizes scale with the number of galaxies in the group, as in Figure \ref{fig:ssfr_compare_targetmass}. The grey squares have both lower limits on SFR and on $\mathrm{A_{V,neb}}$. Nebular attenuation increases along the star-forming sequence, towards the upper-right.}
\label{fig:sfr_mass_avneb}
\end{figure}

\section{Results}
\label{sec:results}

Here we describe the findings from the composite SEDs, \texttt{Prospector} fits, and the stacked spectra. In Section \ref{subsec:photospec_results} we present the SFR, emission line, and UVJ properties, while Section \ref{subsec:dust_results} focuses on dust. A summary of the measured properties of our composite SED groups are presented in Table \ref{tab:sed_table}.

\subsection{Photometric and Spectral Properties}
\label{subsec:photospec_results}

In Figure \ref{fig:composite_seds_page0}, we show an overview of all the composite SEDs and stacked spectra, as well as each group's location in the SFR-mass diagram, the UVJ diagram, and the Baldwin, Phillips, and Terlevich (BPT) diagram \citep{baldwin_classification_1981}. While many individual galaxies do not have strong enough emission detections to make measurements of Balmer decrement-corrected \halpha SFR and metallicity, we can make these measurements from the composite spectra for nearly all groups. In Figure \ref{fig:mass_sfr_uvj_bpt}, we summarize the fundamental galaxy properties as derived from the composite SED and emission-line measurements, displaying how they relate to each other and the individual galaxies for which we can make measurements. 

\begin{figure*}[]
\vglue -5pt
\centering
\includegraphics[width=0.95\textwidth]{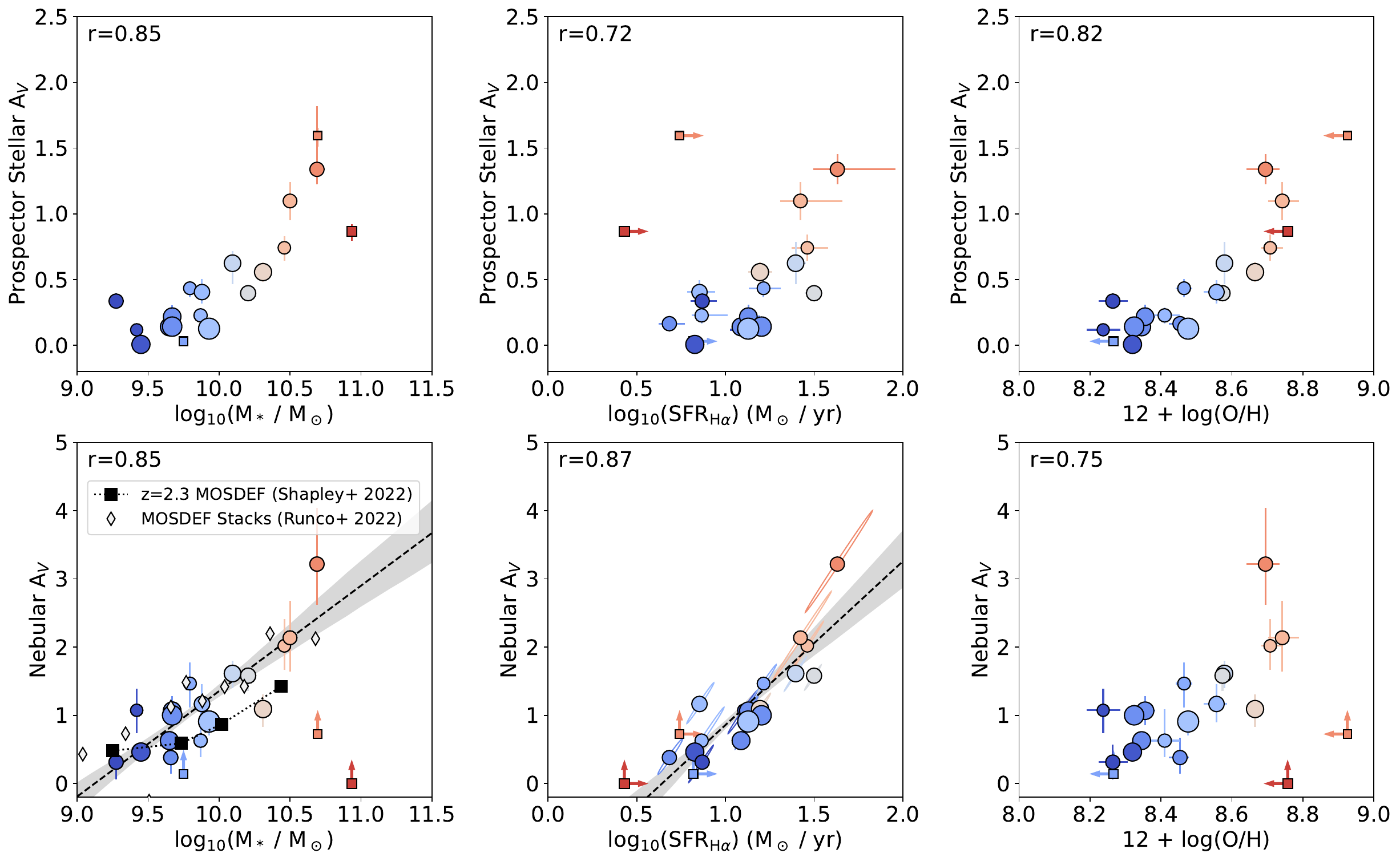}
\caption{
Top row: Stellar \AV (derived from \texttt{Prospector} composite SED fits) vs mass (left), SFR (center), and metallicity (right) for the 20 galaxy groups. Bottom row: Nebular \AV (derived from stacked spectra Balmer decrement) vs mass (left), SFR (center), and metallicity (right). Symbols and colors are the same as in Figure \ref{fig:ssfr_compare_targetmass}. Correlation coefficients are listed in the top-left of each panel. For the nebular \AV vs.\ mass and SFR (lower-left and lower-middle), we also show linear fits (black dashed line) and corresponding bootstrapped one-sigma uncertainties (grey shaded region). For the lower-middle panel, uncertainties are shown as ellipses due to the dependence of SFR on nebular \AV. We also compare our findings with \citet{shapley_mosfire_2022} (black squares) and \citet{runco_mosdef_2022} (gray diamonds) in the lower-left panel. We see a clear trend of stellar \AV increasing with increasing mass, and a potential turnover at high masses. For nebular \AV, we find tight relations with increasing mass and SFR.}
\label{fig:dust_panel}
\end{figure*}

\begin{figure*}[]
\vglue -5pt
\centering
\includegraphics[width=0.95\textwidth]{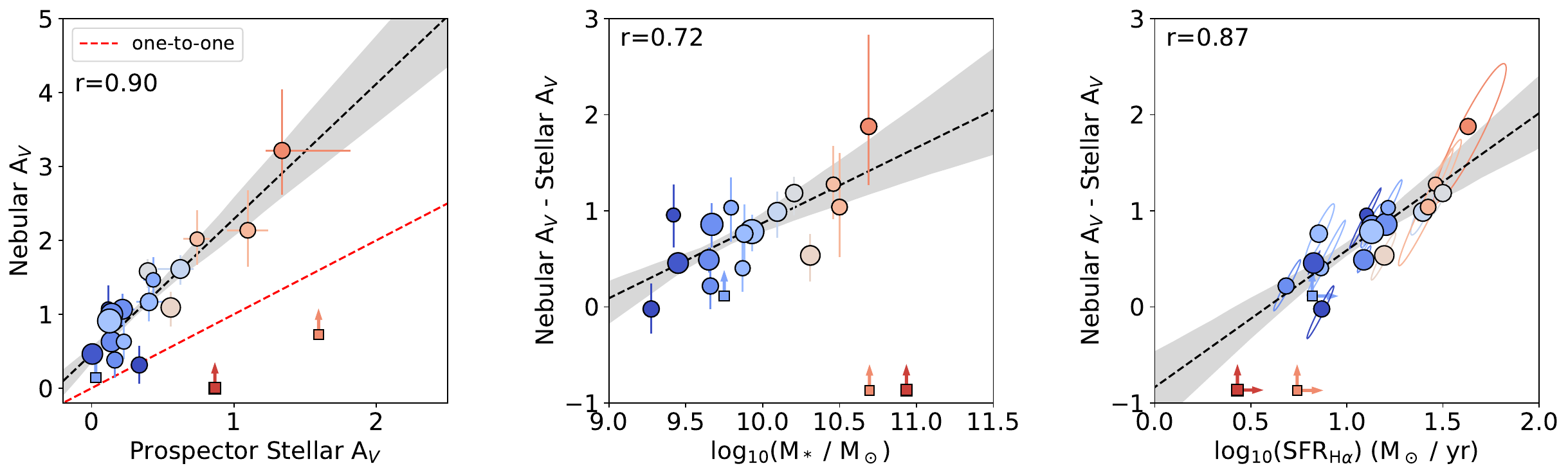}
\caption{
Symbols and colors are the same as in Figure \ref{fig:ssfr_compare_targetmass}. Left: Nebular \AV vs.\ stellar \AV. The properties are well-correlated, with nebular \AV being larger than stellar \AV by roughly a factor of 2 for all groups with \hbeta detections. Middle: Nebular \AV $-$ stellar \AV (\AV excess) vs.\ mass. Right: \AV excess vs.\ SFR. Uncertainties are shown as error ellipses due to the dependence of nebular \AV on SFR. \AV excess is more strongly related with SFR than with mass. This SFR-dependence of \AV excess is in agreement with \citet{reddy_mosdef_2015}. The strong correlation with SFR suggests that nebular attenuation may be dominated by a different component than stellar attenuation, and the relative importance of these components changes with galaxy SFR.
}
\label{fig:av_compare}
\end{figure*}

First, in the top-left panel of Figure \ref{fig:mass_sfr_uvj_bpt}, we examine the UVJ diagram \citep{wuyts_what_2007} for the composite SEDs. Since they are selected into groups by SED shape, galaxies in the same group occupy a very similar spot in the UVJ diagram, as shown in Figure \ref{fig:composite_seds_page0}. We find the expected trend that more massive galaxies tend towards the top-right of the star-forming sequence, with one exception: the most massive group is lower down and offset from the sequence, towards bluer $V-J$. This offset may point towards an evolutionary path along the UVJ diagram, where the most massive galaxies slightly drop down the star-forming sequence before moving to the quiescent side. We will re-visit this evolutionary discussion in Section \ref{sec:discussion}. 

In the top-right panel of Figure \ref{fig:mass_sfr_uvj_bpt}, we display sSFR vs.\ mass using the Balmer decrement-corrected \halpha SFRs and median mass of the galaxies in each group. We also show the data points of the individual MOSDEF galaxies, but caution that 32\% of these individual sSFRs are only computed from \halpha and an SED dust-correction, since \hbeta is undetected (see Section \ref{subsec:mosdef_survey}). The galaxies largely fall in expected positions along the star-forming main sequence (SFMS), tracing the median of the background points. Notably, the slope of the SFMS does not seem to be affected by the fact that prior works may have excluded the most-obscured galaxies due to faint emission lines. We also compare our sample to the \citet{leja_new_2021} star-forming main sequence (ridge) measured from SED fitting of 3D-HST galaxies. Our \halpha SFRs are in good agreement with these SED-based SFRs. From this figure, it is challenging to see if there is a turn-off at high masses or if we are simply underestimating the SFRs of high-mass galaxies, as we only have lower-limits for the composite groups. The individual galaxies do seem to lie below the SFMS at high masses, but many of these points have only \halpha\ detections and therefore may have uncertain dust corrections. We also note that the \halpha SFR may be overestimated for the highest-mass bin (see Section \ref{subsec:prospector}) and that its \texttt{Prospector} SED-based sSFR is $10^{-11.80}\ \rm{M}_\odot \rm{yr}^{-1}$, which is significantly below the star-forming main sequence.

We also plot each group in mass-metallicity space (bottom-right of Figure \ref{fig:mass_sfr_uvj_bpt}, gray points are $3\sigma$ detections in all four lines), using the O3N2 metallicity measure described in section \ref{subsec:metallicity_measure}. Our data points form a tight relation, as expected, with higher mass galaxies showing higher metallicities. Our results are consistent with previous MOSDEF mass-metallicity relations from the MOSDEF $z=1.5$ sample from \citet{topping_mosdef_2021} and MOSDEF $z=2.3$ sample from \citet{sanders_mosdef_2018}. Our sample, with a mean redshift of $z=1.9$, indeed falls between these lines.

Finally, we examine the BPT diagram in the lower-left of Figure \ref{fig:mass_sfr_uvj_bpt}, where the background points are individual galaxies with $3\sigma$ detections in all four lines. We compare our measurements from composite SEDs to individual measurements of MOSDEF galaxies from \citet{shapley_mosdef_2015}, finding strong agreement. The groups follow the expected mass trend, with more massive galaxies towards the lower-right of the diagram. The two most massive groups fall along the sequence, but their lower-limits allow for higher \OIII/\hbeta, which could push them into the regime of non-star-forming ionization. The SFR comparison in Figure \ref{fig:ssfr_compare_targetmass} indicates that, for group 20, the \halpha SFR may be overestimated. Therefore, the \halpha emission in this group may not originate from star formation, but instead from AGN or hot evolved stars.

\subsection{Dust Properties}
\label{subsec:dust_results}

Next, we use the groups to inform a more complete view of dust in $z=2$ galaxies. The dust properties are synthesized from Balmer decrement measurements from the stacked spectra (section \ref{subsec:emission_fit}) as well as outputs from the \texttt{Prospector} modelling of the composite SEDs (section \ref{subsec:prospector}). These measurements are visualized in Figures \ref{fig:sfr_mass_avneb}, \ref{fig:dust_panel} and \ref{fig:av_compare}.

First, we find that nebular \AV, measured from the Balmer decrement, increases along the star-forming sequence up to a maximum of almost 3.5 mag (Figure \ref{fig:sfr_mass_avneb}). This result is in agreement with \citet{runco_mosdef_2022}, which measured Balmer decrements from stacked spectra in bins of stellar mass for a complete sample of MOSDEF galaxies, including those without \hbeta detections. Studies have also shown similar trends for the stellar \AV \citep[e.g.,][]{cullen_vandels_2018}, as well as the dust emission \citep[e.g.,][]{heinis_hermes_2014}. 

In the left panels of Figure \ref{fig:dust_panel}, we compare both stellar and nebular \AV measurements with stellar mass. For both stellar \AV (measured from the composite SEDs) and nebular \AV (from Balmer decrements), we see a clear trend of increasing attenuation with increasing mass. However, the most massive group lies below this trend for both nebular and stellar \AV. For the stellar \AV, we see a possible turnover at the highest mass bin, where the \AV begins to drop. Since we only measure lower-limits, it is unclear if the massive galaxies indeed fall on the nebular \AV trend but the measurements are not sensitive enough to detect it, or instead if the turnover is again present. We will revisit this possibility in Section \ref{sec:discussion}. 

We measure the following relation between the Nebular \AV and stellar mass:
\begin{equation}
    \AVneb = 1.55 \times \log_{10}M_* - 14.10. 
\end{equation}
Interestingly, the relation from the stacked spectra predicts higher nebular \AV\ at a given mass than what was measured in the $z=2.3$ MOSDEF sample of individual star-forming galaxies from \citet{shapley_mosfire_2022}, and a similar relation to what was found in MOSDEF stacked spectra from \citet{runco_mosdef_2022}. A likely explanation for this offset is that the $z=2.3$ MOSDEF sample requires a $3\sigma$ detection of the \hbeta line for individual galaxies, while this work and the stacked spectra measurements from \citet{runco_mosdef_2022} do not. Therefore, there appears to be a bias in nebular \AV measurements when excluding the systems without \hbeta\ detections. 

We also examine the effects of both SFR and metallicity on the nebular and stellar \AV. Following the correlation coefficients in Figure \ref{fig:dust_panel}, we see a moderate relation between stellar \AV and SFR, and a strong relation with metallicity. On the other hand, nebular \AV correlates extremely well with SFR and reasonably well with metallicity. This tight relation of nebular \AV with \halpha SFR may be expected, since the SFR measurement includes a nebular \AV dust correction of the \halpha luminosity. The correlation is reflected in the uncertainty ellipses, which point in a similar direction as the observed trend. We do observe that the range of observed SFRs is much wider than the uncertainties for most of the points, so there is likely an underlying relationship between SFR and nebular \AV despite the artificial tightening. For the observed correlation between nebular \AV and SFR, we measure 
\begin{equation}
    \AVneb = 2.40 \times \log_{10}\left(\frac{\mathrm{SFR}}{\mathrm{M_\odot yr}^{-1}}\right)  - 1.54.
\end{equation}

Next, we compare the nebular \AV to the stellar \AV in Figure \ref{fig:av_compare}. We plot both the linear least-squares best-fit to the points without lower-limits and a one-to-one line. We see clearly that the nebular \AV is larger than the stellar \AV by roughly a factor of 2. To further examine the excess nebular \AV compared to stellar \AV, we compare the excess to other properties. Stellar mass only seems to show a slight trend with the \AV excess. We also compare the \AV excess to SFR, sSFR, and metallicity, and found that the strongest trend is between \AV excess and SFR \citep[also observed in][]{price_direct_2014}. This correlation is in qualitative agreement with \citet{reddy_mosdef_2015} for individual MOSDEF galaxies. The relation observed here has significantly lower scatter and is based on a larger dataset, but covers a smaller range of SFRs due to the composite grouping. The best-fit relation for this correlation is
\begin{equation}
    \AVneb - \AVstellar = 1.83 \times \log_{10}\left(\frac{\mathrm{SFR}}{\mathrm{M_\odot yr}^{-1}}\right) + 0.46
\end{equation}.
We also note that since nebular \AV is used to correct the SFRs, these variables are not completely independent, and the correlated uncertainties are shown as ellipses. The uncertainties are not directly along the relation, and we cover a wide range of galaxy SFRs, so there seems to be a strong intrinsic correlation in addition to effects from the dependence of the variables.

\begin{table*}[t]
  \centering
  \vspace{0.8cm}  
    \caption{\label{tab:sed_table} Summary of properties from the composite group, stacked spectra and \texttt{Prospector} fitting. SFR, metallicity, and $\frac{\halpha}{\hbeta}$ are given as $2\sigma$ limits for groups 8, 19, and 20.}
  \begin{tabular}{cccccccccccc}\hline\hline
    \multicolumn{6}{c}{Composite Group}  & \multicolumn{3}{c}{Stacked Spectra} & \multicolumn{3}{c}{\texttt{Prospector} Fit} \\
    \cmidrule(lr){1-6}
    \cmidrule(lr){7-9}
    \cmidrule(lr){10-12}
    $\rm{Group}$ & $N$ & $\widetilde{z}$ & ${\log(M_*)}$\footnote{median stellar mass of the individual galaxies in each group} & $U-V$ & $V-J$ & $\log(\rm{SFR})$ & $12 + \log{\left(O/H\right)}$\footnote{derived from the O3N2 indicator. See Section \ref{subsec:metallicity_measure}}  &  $\halpha/\hbeta$ & $A_\mathrm{V}$ & $\log(\rm{sSFR})$ & Metallicity \\
    &&& $\mathrm{M}_\odot$ &&& $\mathrm{M}_\odot \rm{yr}^{-1}$ &&&& $\mathrm{M}_\odot \rm{yr}^{-1}$ & $\log(Z/\mathrm{Z}_\odot$) \\
    \cmidrule(lr){1-6}
    \cmidrule(lr){7-9}
    \cmidrule(lr){10-12}
    1 & 30 & $2.22$ & $9.28$ & $0.54$ & $0.25$ & $0.87_{-0.07}^{+0.08}$ & $8.26_{-0.04}^{+0.04}$ & $3.16_{-0.25}^{+0.27}$ & $0.34_{-0.04}^{+0.06}$ & $-8.12_{-0.03}^{+0.04}$ & $-0.32_{-0.05}^{+0.04}$ \\ 
2 & 22 & $1.56$ & $9.42$ & $0.52$ & $0.26$ & $1.10_{-0.08}^{+0.10}$ & $8.24_{-0.05}^{+0.05}$ & $4.03_{-0.41}^{+0.43}$ & $0.12_{-0.03}^{+0.03}$ & $-8.67_{-0.03}^{+0.02}$ & $-0.22_{-0.04}^{+0.04}$ \\ 
3 & 48 & $2.12$ & $9.45$ & $0.57$ & $0.30$ & $0.83_{-0.03}^{+0.03}$ & $8.32_{-0.03}^{+0.02}$ & $3.32_{-0.11}^{+0.11}$ & $0.01_{-0.00}^{+0.01}$ & $-9.36_{-0.02}^{+0.01}$ & $-1.37_{-0.02}^{+0.03}$ \\ 
4 & 47 & $2.29$ & $9.65$ & $0.59$ & $0.22$ & $1.09_{-0.03}^{+0.04}$ & $8.35_{-0.03}^{+0.02}$ & $3.50_{-0.14}^{+0.14}$ & $0.14_{-0.04}^{+0.05}$ & $-8.60_{-0.03}^{+0.04}$ & $-0.10_{-0.04}^{+0.05}$ \\ 
5 & 30 & $1.51$ & $9.66$ & $0.66$ & $0.43$ & $0.68_{-0.06}^{+0.09}$ & $8.45_{-0.03}^{+0.02}$ & $3.23_{-0.24}^{+0.31}$ & $0.16_{-0.04}^{+0.05}$ & $-9.39_{-0.04}^{+0.04}$ & $-1.33_{-0.05}^{+0.04}$ \\ 
6 & 45 & $2.20$ & $9.67$ & $0.60$ & $0.30$ & $1.13_{-0.05}^{+0.07}$ & $8.36_{-0.02}^{+0.03}$ & $4.03_{-0.27}^{+0.28}$ & $0.22_{-0.09}^{+0.09}$ & $-8.70_{-0.06}^{+0.06}$ & $0.06_{-0.06}^{+0.05}$ \\ 
7 & 55 & $2.30$ & $9.67$ & $0.58$ & $0.25$ & $1.20_{-0.04}^{+0.05}$ & $8.32_{-0.02}^{+0.02}$ & $3.94_{-0.17}^{+0.20}$ & $0.14_{-0.04}^{+0.06}$ & $-9.17_{-0.05}^{+0.04}$ & $-1.34_{-0.04}^{+0.05}$ \\ 
8 & 11 & $2.18$ & $9.75$ & $0.79$ & $0.30$ & $>0.82$ & $<8.45$ & $>2.99$ & $0.03_{-0.02}^{+0.02}$ & $-9.58_{-0.06}^{+0.06}$ & $-1.35_{-0.03}^{+0.04}$ \\ 
9 & 24 & $2.23$ & $9.80$ & $0.85$ & $0.32$ & $1.21_{-0.08}^{+0.10}$ & $8.47_{-0.02}^{+0.02}$ & $4.57_{-0.48}^{+0.48}$ & $0.43_{-0.07}^{+0.05}$ & $-9.23_{-0.07}^{+0.06}$ & $-0.12_{-0.11}^{+0.09}$ \\ 
10 & 26 & $2.17$ & $9.87$ & $0.60$ & $0.47$ & $0.87_{-0.06}^{+0.15}$ & $8.41_{-0.05}^{+0.04}$ & $3.50_{-0.26}^{+0.55}$ & $0.23_{-0.06}^{+0.05}$ & $-8.76_{-0.05}^{+0.04}$ & $0.09_{-0.04}^{+0.05}$ \\ 
11 & 35 & $1.53$ & $9.88$ & $0.78$ & $0.69$ & $0.85_{-0.07}^{+0.09}$ & $8.56_{-0.04}^{+0.03}$ & $4.16_{-0.34}^{+0.40}$ & $0.40_{-0.09}^{+0.10}$ & $-9.11_{-0.06}^{+0.06}$ & $0.11_{-0.05}^{+0.04}$ \\ 
12 & 65 & $2.30$ & $9.93$ & $0.70$ & $0.36$ & $1.13_{-0.05}^{+0.05}$ & $8.48_{-0.01}^{+0.02}$ & $3.83_{-0.24}^{+0.21}$ & $0.13_{-0.03}^{+0.05}$ & $-8.98_{-0.03}^{+0.05}$ & $0.02_{-0.05}^{+0.04}$ \\ 
13 & 40 & $2.09$ & $10.10$ & $0.81$ & $0.68$ & $1.40_{-0.05}^{+0.06}$ & $8.58_{-0.01}^{+0.01}$ & $4.79_{-0.32}^{+0.30}$ & $0.62_{-0.16}^{+0.09}$ & $-8.83_{-0.10}^{+0.06}$ & $0.12_{-0.05}^{+0.05}$ \\ 
14 & 34 & $2.29$ & $10.20$ & $0.76$ & $0.60$ & $1.50_{-0.05}^{+0.05}$ & $8.57_{-0.01}^{+0.02}$ & $4.74_{-0.33}^{+0.26}$ & $0.40_{-0.06}^{+0.04}$ & $-8.76_{-0.05}^{+0.04}$ & $0.16_{-0.02}^{+0.02}$ \\ 
15 & 43 & $2.23$ & $10.31$ & $0.92$ & $0.77$ & $1.19_{-0.07}^{+0.07}$ & $8.67_{-0.02}^{+0.02}$ & $4.06_{-0.32}^{+0.28}$ & $0.56_{-0.08}^{+0.07}$ & $-9.05_{-0.05}^{+0.05}$ & $0.12_{-0.04}^{+0.04}$ \\ 
16 & 23 & $2.24$ & $10.46$ & $1.07$ & $0.85$ & $1.46_{-0.09}^{+0.12}$ & $8.71_{-0.03}^{+0.04}$ & $5.46_{-0.58}^{+0.73}$ & $0.74_{-0.10}^{+0.09}$ & $-9.19_{-0.07}^{+0.05}$ & $0.07_{-0.04}^{+0.05}$ \\ 
17 & 28 & $1.54$ & $10.50$ & $1.12$ & $1.04$ & $1.42_{-0.11}^{+0.24}$ & $8.74_{-0.04}^{+0.05}$ & $5.67_{-0.84}^{+1.08}$ & $1.10_{-0.15}^{+0.14}$ & $-9.16_{-0.09}^{+0.09}$ & $0.10_{-0.05}^{+0.05}$ \\ 
18 & 30 & $1.63$ & $10.69$ & $1.40$ & $1.34$ & $1.63_{-0.14}^{+0.33}$ & $8.69_{-0.05}^{+0.04}$ & $8.01_{-1.39}^{+2.43}$ & $1.34_{-0.12}^{+0.48}$ & $-9.50_{-0.07}^{+0.06}$ & $0.00_{-1.27}^{+0.07}$ \\ 
19 & 10 & $1.51$ & $10.70$ & $1.62$ & $1.66$ & $>0.74$ & $<9.10$ & $>3.61$ & $1.59_{-0.08}^{+0.06}$ & $-9.43_{-0.06}^{+0.04}$ & $0.14_{-0.04}^{+0.03}$ \\ 
20 & 14 & $2.19$ & $10.93$ & $1.46$ & $1.06$ & $>0.43$ & $<9.03$ & $>2.86$ & $0.87_{-0.07}^{+0.06}$ & $-11.80_{-0.31}^{+0.14}$ & $-1.28_{-0.08}^{+0.14}$ \\ 
\hline
\end{tabular}
\vspace{-0.8cm}
\end{table*}

\section{Discussion} \label{sec:discussion}

Given that we measure galaxy properties across a wide range of SED types, we now examine how the results provide insights into the evolution of $z\approx2$ galaxies (Section \ref{subsec:evolution}). We also discuss what the dust attenuation results imply about the dust geometry in $z\approx2$ galaxies (Section \ref{subsec:dust_model}). Finally, we present caveats to this work (Section \ref{subsec:caveats}).

\subsection{Implications for Galaxy Evolution} \label{subsec:evolution}

Due to the wide range of masses and SFRs encompassed in our sample, our galaxy groups represent different evolutionary stages. In particular, we have highly sampled SEDs from which we can derive more robust ensemble properties than for individual galaxies, including better measurements of specific SFR and stellar \AV. We also have spectroscopic properties from stacks, which can place stronger limits on the nebular attenuation and \halpha SFRs of the highest mass galaxies. Often, these high-mass galaxies lack significant Balmer emission, but they can still be measured through stacking. 

We do note, however, that these groups cannot form an evolutionary sequence since they are all from a similar redshift regime. However, the groups can capture galaxies at different evolutionary stages --- perhaps one group is quenching, while another is undergoing an early starburst. The location on the groups on various diagrams can give insights into evolution. In particular, group 20 has a number of signs that indicate it may contain quenching galaxies, including the high mass, the location on UVJ and BPT diagrams, and the low SED-based SFR (see Section \ref{subsec:prospector}).

In UVJ space (top-left of Figure \ref{fig:mass_sfr_uvj_bpt}), we recover the trend that stellar mass increases towards the upper right of the diagram \citep{williams_evolving_2010, patel_uvj_2012}. We also see from the other panels of Figure \ref{fig:mass_sfr_uvj_bpt} that as galaxies increase in mass, their metallicity increases and sSFR decreases. However, as we trace this trend in the UVJ diagram, we notice that the highest mass group (20) is offset towards lower in $V-J$ than expected, potentially indicating evolutionary movement. As galaxies grow in stellar mass and their sSFR drops, they may fall back down the star-forming sequence of the UVJ diagram before climbing towards the quiescent region. We therefore caution that the UVJ colors alone may not be sufficient to classify galaxies --- the same space of the UVJ diagram may be occupied by moderate-mass galaxies that are evolving towards the top-right and higher-mass, lower-sSFR galaxies that are moving towards the quiescent box. These trends can be seen in some of the simulated galaxies in \citet{akins_quenching_2022}, and spectral stacks of MOSDEF galaxies in this transitional region of the UVJ diagram have been previously studied by \citet{zick_mosdef_2018}. Instead, a galaxy's place on the SFR-mass diagram may be more closely linked to its evolutionary stage than its location on the UVJ diagram.

Along with the decrease in sSFR observed in group 20, we find that the stellar \AV decreases and that lower-limits allow for decreases in nebular attenuation as well. These results support the finding that high-mass galaxies are shutting down star formation (leading to lower nebular \AV) and are perhaps expelling or destroying ISM dust (leading to the lower stellar \AV). 

Finally, we speculate on the nature of this evolution from the BPT diagram. For group 20, the \halpha SFR is especially higher than the \texttt{Prospector} SED-based SFR. Therefore, it is possible that some \halpha emission may originate from AGN. AGN feedback could be one of the mechanisms that cause this group to shut off star formation and dispel ISM dust, producing all of the other observed trends. Alternatively, the additional \halpha may originate from hot evolved stars, which become more common as the population ages and may also contribute to the observed trends \citep[e.g.,][]{cid_fernandes_alternative_2010}.

\subsection{Dust Attenuation Implications} \label{subsec:dust_model}


By examining the results of the attenuation measurements, we gain new insights into the dust geometry at cosmic noon. The following results place constraints on the possible dust geometries.

We first refer to the result from \citet{lorenz_updated_2023}, finding that MOSDEF $z=2$ galaxies do not show changes in nebular \AV or stellar \AV with inclination angle. Consequently, they proposed that the ISM dust does not play a large role in the variation of stellar or nebular attenuation, and instead the attenuation is likely local to star-forming regions.

We recover the well-known trend that both stellar and nebular attenuation are strongly correlated with galaxy mass \citep[e.g.,][]{garn_predicting_2010, dominguez_dust_2013, price_direct_2014, whitaker_constant_2017, cullen_vandels_2018, shapley_mosfire_2022, runco_mosdef_2022, maheson_unravelling_2024}. We also show that nebular \AV has a very tight correlation with SFR, which is partly due to the dependence of SFR on nebular \AV. On the other hand, the stellar attenuation primarily correlates with mass and does not show as strong of a relation with SFR. The connection between \AV and mass can be explained by more massive galaxies having formed more stars over their lifetime and having larger gravitational potentials. Since more stars have been formed, they have released more metals and dust into the ISM. Since they also have larger potential well, outflows are less efficient at removing the metals and dust. The higher-metallicity gas leads to a higher dust-to-gas ratio \citep{popping_dust--gas_2022} and thus higher dust column density in new star-forming regions, which is where the bulk of the attenuation takes place.

We find that the nebular \AV is larger than the stellar attenuation for all groups with measured Balmer decrements. Additionally, the excess of nebular attenuation compared to the stellar attenuation correlates with galaxy mass and strongly with SFR \citep[Figure \ref{fig:av_compare}, and see][]{reddy_mosdef_2015}. The extra attenuation may indicate that star formation occurs in both typical and heavily-obscured star-forming regions, with the relative importance of these components depending on the SFR of the galaxy. At high SFR, a significant fraction of the star formation would occur in the highly-obscured regions, producing strong nebular attenuation without much contribution to the stellar \AV. Thus, we would observe excess nebular attenuation relative to the stellar \AV. At low SFR, there would not be substantial heavily-obscured star formation, and so the nebular \AV is produced by mostly the same material as regions producing stellar \AV, causing the excess to trend towards 0. From our findings it is unclear whether these highly-obscured star-forming regions take the form of star-forming clumps that have been observed in $z=2$ galaxies, central star-formation, or some other source. Similar pictures with areas of increased obscuration are proposed in \citet{reddy_mosdef_2015} and \citet{lorenz_updated_2023}.

Our results are consistent with the finding that galaxies at $z=2$ appear to have clumpy star formation \citep{wuyts_smoother_2012, schreiber_constraints_2011}, and higher star-forming galaxies may have more star formation occurring in clumps. Using low-redshift starburst galaxies, \citet{hinojosa-goni_starburst_2016} found that clump mass and SFR scale with their host galaxy. Therefore, it would be unsurprising for the higher mass and SFR galaxies in MOSDEF to have large clumps with highly obscured star formation, and therefore a source of excess nebular attenuation. 

On the other hand, many galaxies at $z=2$ have been observed to have highly-obscured central star-forming regions \citep[e.g.,][]{nelson_where_2016, chen_extended_2020}. The highly-obscured central star formation that we predict here could also be in central regions, but the MOSDEF galaxies are unlikely to have completely obscured central regions due to their SED SFRs (with FIR included) being similar to their \halpha SFRs \citep{shivaei_mosdef_2016}. In another study, \citet{fetherolf_mosdef_2021} found that the \halpha-to-UV SFR ratio is higher in the central regions of MOSDEF galaxies, and further increases with increasing SFR. This result indicates that the centers of MOSDEF galaxies could host the highly-obscured star-forming regions.

To distinguish between the scenarios of clumpy or central star-formation, we require spatially resolved star formation and dust maps of $z=2$ galaxies. With JWST, tracing the structure and location of dust in distant galaxies may now be possible.

\subsection{Caveats}
\label{subsec:caveats}

While our grouping and stacking methods enable analysis of a more complete sample of galaxies than is typically possible, there are some limitations to our methods. Here we discuss a few of the most notable considerations.

First, we are inferring a number of properties from fits to stacked spectra and composite SEDs. Most concerns about stacking arise from combining different stellar populations. However, here the groups were formed from galaxies which have similar SED shapes, and thus are likely to have similar underlying stellar populations, so we do not expect stacking to strongly bias our results. We also note that we are stacking the spectral luminosities, and are not normalizing the spectra by the \halpha luminosity in order to maximize the signal-to-noise ratio of the stacked spectra. With our method, we are assuming that galaxies with similar SED shapes have similar properties of Balmer decrements, sSFR, and metallicities, so our measurements should not be affected by this decision. 

Second, we recognize that all of the galaxies are at different redshifts. Many of the groups have similar redshift distributions, dominated by galaxies in the higher redshift regime (see Table \ref{tab:sed_table} to find median redshifts). Six of the groups (2, 5, 11, 17, 18, 19) have a larger fraction of their galaxies from the lower redshift regime. These groups are spread across the evolutionary sequence, so we do not expect any large biases. However, there may be still be a small bias in the properties that we measure when comparing across the groups at different redshifts.

Third, despite our best efforts to capture a complete sample of star-forming galaxies at $1.37\leq z \leq 2.61$, we may still be missing the most obscured systems \citep[e.g.,][]{casey_dusty_2014}. We have made the improvement of including galaxies that do not have significant \halpha and \hbeta detections in an emission-line study, which are typically too faint to include. However, anything that is completely obscured will not have strong enough emission lines to make a spectroscopic redshift measurement, and thus is not included in the stacks. Furthermore, for very dusty galaxies, spectral breaks are not visible and thus photometric redshifts are less certain, so these dusty galaxies may have been missed by the MOSDEF survey. Fortunately, the primary results of this paper reflect the dust properties of moderately-attenuated star-forming galaxies, so missing the most dusty systems is not concerning.

Finally, we note that the stellar \AV may still be biased due to our modeling procedure. We re-ran the \texttt{Prospector} fits with metallicity fixed at half-solar, and the results did not change. Therefore, even though the \texttt{Prospector} SED fits do not constrain metallicity very well, we are not concerned with the accuracy of determining a stellar \AV. However, we lack long-wavelength dust emission for most of these galaxies, and therefore some uncertainties with the dust properties remain. SED fitting to MOSDEF galaxies with longer wavelength coverage is only possible for a small sub-sample, either with bright Herschel flux or follow-up ALMA imaging, and can be found in \citet{shivaei_mosdef_2016, shivaei_infrared_2022}.

\section{Summary} \label{sec:summary}

In this work, we examine 660 galaxies at $1.37\leq z\leq 2.61$ from the spectroscopic MOSDEF survey by algorithmically dividing them into groups with similar SED shapes. By combining the individual photometric measurements in each group, we form a highly-sampled composite SED. We also stack the individual MOSDEF spectra of all of the galaxies in each group. These SEDs and stacked spectra enable measurements of emission lines and aggregate dust and stellar population properties that are not possible for all individual galaxies. In particular, 196 ($32\%$) of the individual galaxies with \halpha detections do not have detected \hbeta, but are still able to contribute to the stacks. Furthermore, since the different groups represent different evolutionary stages, we can unravel how the emission-line and SED properties of galaxies change with evolutionary stage. We summarize our key findings:

\begin{itemize}
    \item The stacking technique employed in creating composite SEDs has lead to a more complete sample of galaxies --- we are able to include an additional 253 galaxies in the stacked spectra that do not have detected \hbeta lines. Our method is complementary to the work by \citet{runco_mosdef_2022}, who stack the MOSDEF galaxies in bins of stellar mass. 
    \item The 20 galaxy groups formed in this work have a wide range of properties, allowing measurements at different galaxy evolutionary stages. We recover well-known trends for mass-SFR, mass-metallicity, and the star-forming sequence in the UVJ and BPT diagrams with low scatter. Thus, the fact that previous studies were not fully complete did not lead to biased trends in these figures, even if they may have missed the most dusty galaxies. However, we do find higher nebular \AV values in the stacks with all galaxies compared to stacks that required the detection of both the \halpha and \hbeta emission lines. Therefore, excluding galaxies with non-detections in the Balmer lines may lead to bias when computing nebular \AV.
    \item \texttt{Prospector} SED fitting seems to reasonably predict \halpha SFRs, both from fits to our own composite SEDs as well as agreement with the work by \citet{leja_new_2021} at $z\approx 2$. The only exception is the highest-mass group, for which the lower limit on the \halpha SFR is significantly higher than the SED SFR. This offset could indicate that the line emission no longer originates from star-forming regions. 
    \item Our groups span a wide range of galaxy masses, SFRs, and locations on the UVJ diagram, indicating that the composite SEDs represent multiple stages of galaxy evolution. The group with the highest mass and lowest sSFR appears to be moving towards the quiescent region of the UVJ diagram, and is lower than would be expected along the diagram's star-forming sequence given its high mass. This group also has lower stellar and possibly lower nebular dust attenuation. The position of this group on the UVJ diagram could be explained by expelling or destroying ISM dust as it quenches its star formation. 
    \item Stellar \AV is well-correlated with stellar mass and metallicity, with the exception of the highest-mass bin, in which the sSFR and stellar \AV are already suppressed. Nebular \AV is well-correlated with stellar mass and SFR. Taken together with the assumption that ISM attenuation is insignificant \citep{lorenz_updated_2023}, these results suggest that individual star-forming regions become more dusty on average as galaxy mass increases.
    \item Nebular \AV is larger than stellar \AV for all groups with $3\sigma$ detections, and this excess correlates most strongly with SFR. We explain this excess with a progressively larger fraction of the star formation occurring in highly-obscured regions with increasing SFR \citep[see also][]{reddy_mosdef_2015}. In higher SFR galaxies, the nebular emission primarily originates from these highly-obscured star-forming regions, while the stellar attenuation is dominated by the star formation occurring in less-obscured regions. In lower SFR galaxies, this highly-obscured star formation is rare, so both the nebular and stellar attenuation originate from the same regions and thus are similar. This highly-obscured star formation may take place in large clumps or galaxy centers.
\end{itemize} 
With current data, it is unclear if the highly-obscured star-forming regions reside in large, dusty clumps or are obscured central star formation. Fortunately, JWST provides a promising avenue to explore further details of the dust geometry at high redshift. First, the added sensitivity and IR measurements can measure more precise emission line decrements in individual galaxies, reducing the need to rely on stacking \citep[e.g.,][]{shapley_jwstnirspec_2023, sanders_direct_2024}. Assessing systems individually allows for a more accurate picture, with fewer assumptions about grouping similar galaxies. Second, the increased observable wavelength regime of JWST allows for access to the Paschen lines, which are excellent tracers of SFR that are less affected by dust \citep[e.g.,][]{reddy_paschen-line_2023, neufeld_fresco_2024}. Finally, JWST's spatial resolution will allow for more precise observations of the location of dust in galaxies \citep[e.g.,][]{bezanson_jwst_2022, rigby_jwst_2023, suess_medium_2024}. By studying how dust is spread throughout $z=2$ systems, we can distinguish between the geometric possibilities in this work and further our understanding of the behavior of dust in the high-redshift universe. 

\color{black}
\begin{acknowledgments}
We thank the referee for a thorough and insightful report that has strengthened this work. BL thanks the insightful comments shared at group meetings, especially with Justus Gibson, Chlo{\"e} Benton, Abby Hartley, Aliza Beverage, Chloe Cheng, and Martje Slob. This research is based on observations made with the NASA/ESA Hubble Space Telescope obtained from the Space Telescope Science Institute, which is operated by the Association of Universities for Research in Astronomy, Inc., under NASA contract NAS 5–26555. These observations are associated with program HST-AR-16141.001-A. This material is based upon work supported by the National Science Foundation Graduate Research Fellowship under Grant No. DGE 2146752. We also acknowledge support from NSF AAG grant Nos. AST- 1312780, 1312547, 1312764, 1313171, 2009313, and 2009085, grant No. AR-13907 from the Space Telescope Science Institute, and grant No. NNX16AF54G from the NASA ADAP program. This work was performed in part at the Aspen Center for Physics, which is supported by National Science Foundation grant PHY-2210452. We wish to recognize and acknowledge the very significant cultural role and reverence that the summit of Maunakea has always had within the indigenous Hawaiian community. We are most fortunate to have the opportunity to conduct observations from this mountain.
\end{acknowledgments}

\vspace{5mm}
\software{\texttt{Prospector} \citep{leja_deriving_2017, johnson_stellar_2021}, \texttt{scikit-learn} \citep{pedregosa_scikit-learn_2011}}
\facilities{HST(STIS), Keck}

\typeout{}\bibliography{MOSDEF}{}
\bibliographystyle{aasjournal}

\end{document}